\documentclass[aps,twocolumn,superscriptaddress,floatfix,longbibliography]
{revtex4-1}

\usepackage{txfonts}
\usepackage{microtype}
\usepackage{ulem}
\usepackage{epsfig}
\usepackage{amssymb}
\usepackage{bm}
\usepackage[dvipsnames]{xcolor}
\usepackage{graphicx}
\usepackage{color}
\usepackage{hyperref}
\usepackage{siunitx}

\usepackage{lipsum}

\begin{document}

\author{J. Maurer}\email{jocmaure@phys.ethz.ch}\address{Department of Physics, ETH Zurich, 8093, Zurich, Switzerland}
\author{B. Willenberg}\address{Department of Physics, ETH Zurich, 8093, Zurich, Switzerland}
\author{B. W. Mayer}\address{Department of Physics, ETH Zurich, 8093, Zurich, Switzerland}
\author{C. R. Phillips}\address{Department of Physics, ETH Zurich, 8093, Zurich, Switzerland}
\author{L. Gallmann}\address{Department of Physics, ETH Zurich, 8093, Zurich, Switzerland}
\author{J. Dan\v{e}k}\address{Max-Planck-Institut f\"ur Kernphysik, Saupfercheckweg 1, 69117 Heidelberg, Germany}
\author{M. Klaiber}\address{Max-Planck-Institut f\"ur Kernphysik, Saupfercheckweg 1, 69117 Heidelberg, Germany}
\author{K. Z. Hatsagortsyan}\email{k.hatsagortsyan@mpi-k.de}\address{Max-Planck-Institut f\"ur Kernphysik, Saupfercheckweg 1, 69117 Heidelberg, Germany}
\author{C. H. Keitel}\address{Max-Planck-Institut f\"ur Kernphysik, Saupfercheckweg 1, 69117 Heidelberg, Germany}
\author{U. Keller}\address{Department of Physics, ETH Zurich, 8093, Zurich, Switzerland}

\title{Interplay between Coulomb-focusing and non-dipole effects in strong-field ionization with elliptical polarization  }


\begin{abstract}
Strong-field ionization and rescattering beyond the long-wavelength limit of the dipole approximation is studied with elliptically polarized mid-IR pulses. We have measured the full three-dimensional photoelectron momentum distributions (3D PMDs) with velocity map imaging and tomographic reconstruction. The ellipticity-dependent 3D-PMD measurements revealed an unexpected sharp, thin line-shaped ridge structure in the polarization plane for low momentum photoelectrons. With classical trajectory Monte Carlo (CTMC) simulations and analytical methods we identified the associated ionization dynamics for this sharp ridge to be due to Coulomb focusing of slow recollisions of electrons with a momentum approaching zero. This ridge is another example of the many different ways how the Coulomb field of the parent ion influences the different parts of the momentum space of the ionized electron wave packet. Building on this new understanding of the PMD, we extend our studies on the role played by the magnetic field component of the laser beam when operating beyond the long-wavelength limit of the dipole approximation. In this regime, we find that the PMD exhibits an ellipticity-dependent asymmetry along the beam propagation direction: the peak of the projection of the PMD onto the beam propagation axis is shifted from negative to positive values with increasing ellipticity. This turnover occurs rapidly once the ellipticity exceeds $\sim$0.1. We identify the sharp, thin line-shaped ridge structure in the polarization plane as the origin of the ellipticity-dependent PMD asymmetry in the beam propagation direction. These results yield fundamental insights into strong-field ionization processes, and should increase the precision of the emerging applications relying on this technique, including time-resolved holography and molecular imaging.
\end{abstract}


\maketitle

\section{Introduction}
\label{sec:intro}
Recently, strong-field ionization in mid-infrared (mid-IR) laser fields has gained a lot of attention for the generation of coherent soft x-rays with high harmonic generation (HHG)  \cite{Popmintchev_2012} and for the discovery of a variety of strong field characteristica, like the observation of holographic electron interferences \cite{Huismans_2011} and low-energy structures \citep{Blaga2009strong, Quan_2009,Wolter_2015x}. These processes are typically described through the recollision of the electron wave packet with the residual ion (or parent ion).
In a two-step model \cite{SimpleManModel},
the electron is released to the continuum and subsequently driven back by the laser field towards the ion core where it can recollide. 
Upon return, the electron can either recombine, scatter inelastically or scatter elastically \cite{pfeifer_2006}. 
Recombination of the electron with the ion leads to HHG \cite{McPherson_1987, Ferray_1988}. Inelastic scattering leads to non-sequential double ionization \cite{Fittinghoff_1992, Walker_1994} and excitation of the ion \cite{Feuerstein_2001}.

However, the vast majority of the rescattering electrons undergoes elastic forward scattering, where the absolute value of momentum of the electron does not change on its trajectory past the parent ion. 
Interference of electrons with different ionization paths that end up with the same final momentum leads to many different structures, such as equidistant peaks in energy from above threshold ionization \cite{Agostini_1979}, electron-diffraction patterns \cite{Blaga_2012}, and holographic interference structures \cite{Huismans_2011}. 
Elastic rescattering can lead to signatures in photoelectron momentum distributions (PMDs) such as Coulomb-focusing \cite{Brabec_1996,Comtois_2005} and low energy structures \cite{Blaga2009strong, Quan_2009,Wolter_2015x}. These kind of signatures in PMDs have been assigned to forward scattering which is most pronounced at slow recollisions, i.e. recollisions where the momentum of the electron approaches zero at the time of recollision \cite{Sasaki_2009, Liu_2010,Yan_2010,Huang_2010, Liu_2011, Kastner_2012,Lemell_2012,Moller_2014,Kelvich_2016}. 

Rescattering effects are the strongest with linear polarization of the laser pulse, however, rescattering is still observed with elliptical polarization, which is possible due to the spread of the returning wave packet \cite{Wang_2009,Mauger_2010,Liu_2012,Shafir_2013,Li_2013}. 
The displacement amplitudes within the simple man's model 
\cite{SimpleManModel}, i.e. $ E_0/\omega^2$ along the long axis and $\epsilon E_0/\omega^2$ along the short axis of the polarization ellipse, depend linearly on the peak electric field $E_0$ and the inverse square of the laser frequency $\omega$.
Therefore, rescattering effects are expected to become more significant for mid-IR wavelengths even at moderate laser intensities.  
Moving from near-IR to mid-IR wavelengths will strongly increase the maximal kinetic energy which allows for diffraction experiments with increased resolution \cite{pullen_2015} and HHG with photon energies up to the order of \mbox{1 keV} \cite{Popmintchev_2012}. 
Furthermore, discoveries like low energy structures \cite{Blaga2009strong, Quan_2009,Wolter_2015x} and holographic interference patterns \cite{Huismans_2011} were first observed at mid-IR wavelengths. 

\begin{figure}[t]
\includegraphics{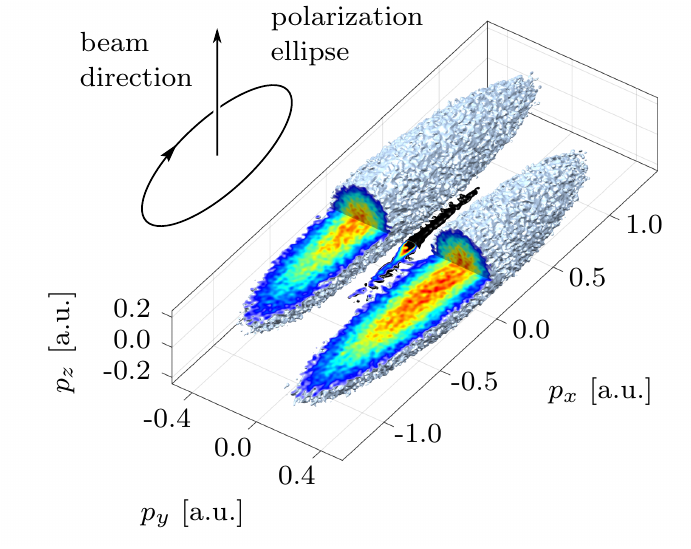}
\caption{Isosurface of a three-dimensional photoelectron momentum
distribution (3D PMD) with a partial cut in the polarization plane $(p_x,p_y)$. The 3D-PMD is recorded using a
mid-IR laser  with  a center wavelength of 3.4 $\mu$m, a pulse length of 50 fs, a peak intensity of $6 \cdot 10^{13}$ W/cm$^2$ and an ellipticity of $\epsilon = 0.11$. The sharp thin line-shaped ridge structure around $p_y = 0$ is clearly separated from the lobes of direct electrons.} 
\label{fig:Surface}
\end{figure}

Another characteristic phenomenon of strong-field ionization at mid-IR wavelengths is the onset of non-dipole effects. When driven by mid-IR laser pulses, the magnetic field induced Lorentz force
starts to become significant for the electron dynamics during the
ionization process. The occurrence of magnetic field effects at long wavelengths has been theoretically predicted  and described as the long-wavelength limit of the dipole approximation \cite{Keitel_1995,Walser_2000a,Milosevic_2000,Kylstra_2001,Reiss_2008,Kohler_2012b}. 
The Lorentz force along the laser propagation direction is responsible for the photon's momentum transfer to electrons observed in \cite{Smeenk_2009momentum}, and for the  momentum partitioning between the ion and the electron during ionization \cite{Klaiber_2013c,Chelkowski_2014,Cricchio_2015,Chelkowski_2015}. 
The laser magnetic field induced drift is known to suppress the recollision and HHG at high laser intensities ($I\gtrsim 10^{17}$ W/cm$^2$ at a laser wavelength $\lambda=800$ nm) \cite{Keitel_1995,Walser_2000a,Milosevic_2000,Kylstra_2001,Dammasch_2001,Klaiber_2005,DiChiara_2008,Ekanayake_2013}, at which the recolliding electron is deflected at the parent ion by more than the electron wave packet size \cite{Walker_2006}. 
At lower laser intensities the recollisions are still possible
because Coulomb focusing can compensate the magnetic-field
induced drift for linear polarization  \cite{Foerre_2006}.

Strong-field ionization experiments at mid-IR wavelengths revealed another phenomenon beyond the long wavelength limit of the dipole approximation \cite{Ludwig_2014} which is relevant within this paper: an initially surprising shift of the peak of the projection of the PMD onto the beam propagation axis opposite to the beam propagation direction was observed. Also, the magnitude of this momentum shift was practically independent of the parent ion. The magnetic-field-induced lateral displacement of the electron and the successive recollision with the Coulomb potential was identified as the cause for this effect. This initial experiment triggered further experimental and theoretical investigations which is the focus of this paper.

\begin{figure}[t]
\includegraphics{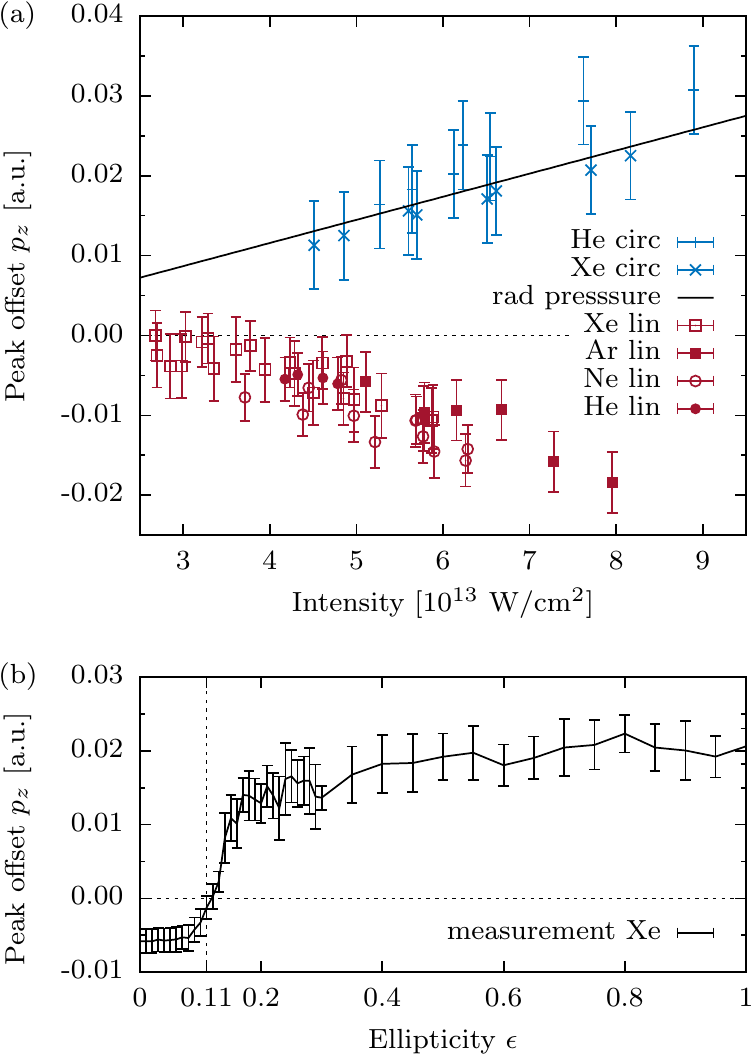}
\caption{ Offsets of the peak of the PMDs projected onto the beam propagation axis ($p_z$) using a mid-IR laser (a) Measurement for circular polarization together with the data for linear polarization taken from Ref. \cite{Ludwig_2014}. We compare our data for circular polarization with the radiation pressure picture used in Ref. \cite{Smeenk_2011} (b)  Measurements as a function of ellipticity for a peak intensity of $6 \cdot 10^{13}$ W/cm$^2$. We observe a transition from negative to positive values of $p_z$ with the zero crossing at $\epsilon \approx$ 0.12. The vertical dashed line indicates the ellipticity corresponding to the 3D PMD from Fig. \ref{fig:Surface}.}
\label{fig:ellShift}
\end{figure} 

Here, we therefore present a more detailed study of ellipticity-dependent strong-field ionization at mid-IR wavelengths taking into account non-dipole effects. We discuss two main results:
 
1) We observed the creation of a sharp, thin line-shaped ridge structure of low-momentum electrons in the polarization plane at small ellipticities (Fig. \ref{fig:Surface}). Our analytical and numerical model show that
this ridge structure stems mainly from Coulomb-focused electrons
undergoing multiple revisits of the parent ion with at least one
significant rescattering event. 

2) We investigated how non-dipole signatures on the PMDs depend on laser ellipticity. 
The shift of the peak of PMDs with respect to the beam propagation axis (Fig. \ref{fig:ellShift})is opposite to the beam propagation direction for both linear polarization and small ellipticites, and is directly related to the sharp, thin line-shaped ridge structure created by Coulomb-focusing in the 3D PMDs. 
With increasing ellipticity we then observe a shift of the PMD peak into beam propagation direction  (Fig.\ref{fig:ellShift} (b)) which gives a direct link between the results observed in Ref. \cite{Ludwig_2014} and the radiation pressure picture in Ref. \cite{Smeenk_2011} that was employed to explain a shift in beam propagation direction. 

The relationship between Coulomb focusing and non-dipole effects explored here increases our understanding of electron ionization dynamics in mid-IR laser fields, helps to better understand initially unexpected features of the observed PMDs, and will allow us to further exploit these effects to significantly enhance the resolution of the attoclock, time-resolved holography and strong-field molecular imaging using mid-IR lasers.
 
The paper is organized as follows: In section \ref{sec:exp}, we present the details of the experiment; in section \ref{sec:PMD3D} we present 3D PMDs recorded with elliptical polarization. We compare the experimental results with classical trajectory Monte-Carlo (CTMC)-simulations; In section \ref{sec:AnaOne} and \ref{sec:AnaTwo}
 we present theoretical models to explain the creation of ridge caused by Coulomb-focused electrons.  The results of the ellipticity-dependent non-dipole effects on the PMD are presented in section \ref{sec:noDipole}.          

\section{Experimental Details}
\label{sec:exp}

PMDs were recorded with a velocity map imaging spectrometer (VMIS) \cite{Eppink_1997, Parker_1997} with the gas nozzle integrated into the repeller to achieve high gas target densities in the interaction region \cite{Ghafur_2009, Weger_2013}. The target was ionized by an optical parametric chirped-pulse amplifier (OPCPA) system based on chirped quasi-phase-matching devices. This system can deliver pulses with duration of \mbox{44 fs} and a pulse energy of \mbox{22 $\mu$J} at a center wavelength of \mbox{3.4 $\mu$m} and a high repetition rate of \mbox{50 kHz} \cite{Mayer_2013, Mayer_2014OSA}. The pulses were focused with a backfocusing dielectric mirror with a focal length of 15 mm into the interaction region. 
The polarization of the laser beam was controlled by two custom-made achromatic MgF$_2$ wave plates. A quarter-wave plate induces the ellipticity and the subsequent half-wave plate controls the orientation of the polarization ellipse. The wave plates were fully characterized via polarimetry measurements where the power transmitted through a polarizer was recorded as a function of the angle between the major polarization axis and the polarizer axis. The polarization state at the desired orientation was extracted via a fit and interpolation of the measured values.

The intensity in all experiments was calibrated with reference measurements at close-to-circular polarization. The radial maximum of the torus-shaped momentum distribution was compared with semiclassical Monte-Carlo simulations \cite{Pfeiffer_2012}.   

Throughout the article, the following coordinate system will be used:  The coordinate $z$ denotes the direction of beam propagation, $x$ the major and $y$ the minor axis of the polarization ellipse and $p_x,p_y,p_z$ the respective electron momenta. $W(p_x,p_y,p_z)$ denotes the PMD, i.e. the amplitude of the photoelectron signal.

The experiments require an accurate determination of the zero momentum spot, in particular on the beam propagation axis. This spot was identified via a sharp point in the center of the PMD recorded with linear polarization that stems from the ionization of atoms that were left in a Rydberg state by the laser pulse and were subsequently ionized by the static electric field of the spectrometer \cite{Nubbemeyer_2008, Smeenk_2011}.  
As these electrons do not interact with the pulse, they are guided by the static electric spectrometer field to the position on the detector that corresponds to zero momentum in the ($p_x,p_z$)-plane. The exact position of zero momentum in $p_z$-direction was determined from the projection of a small range of 0.05 a.u. in $p_x$ of the PMD onto the $p_z$-axis. This projection was fitted with a Lorentzian profile. This method was also applied to find the center in $p_y$-direction.
Throughout the article, atomic units (a.u.) are used.

\section{Three Dimensional Photoelectron Momentum Distributions (3D PMDs)}
\label{sec:PMD3D}

\begin{figure*}
    \includegraphics{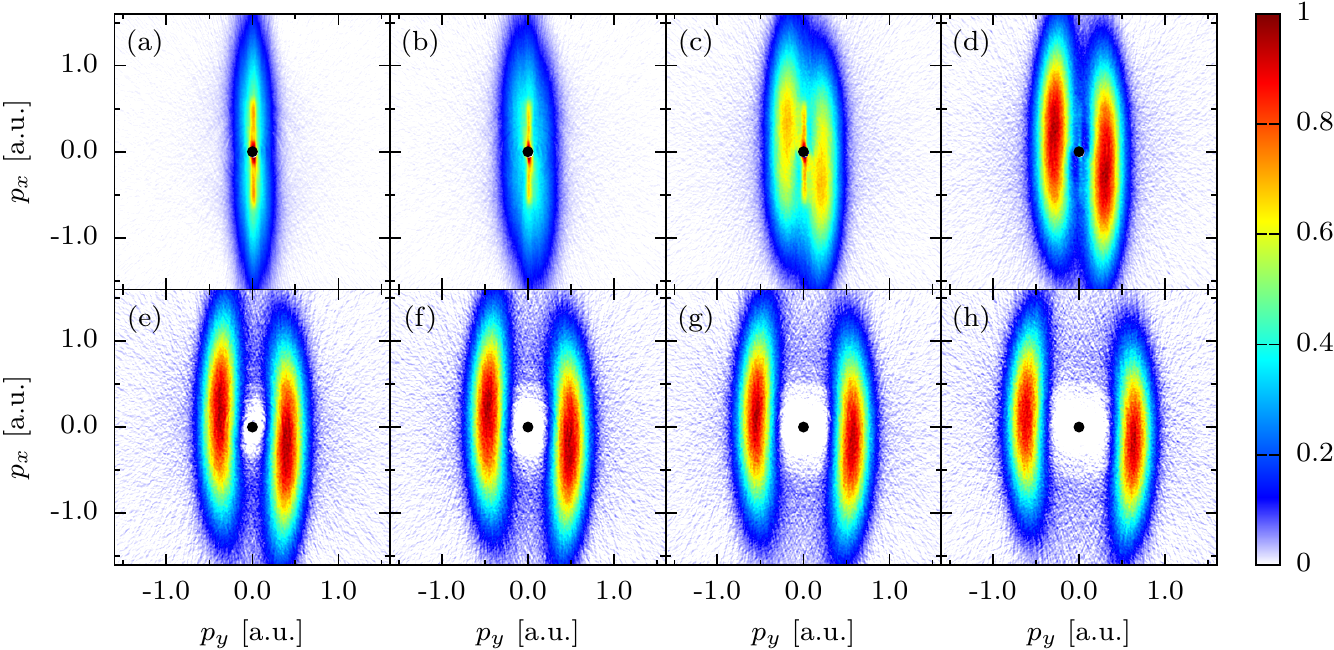}
	\caption{Measured PMDs in the polarization plane measured at a peak intensity of $6 \cdot 10^{13}$ W/cm$^2$ for the  ellipticites 0.0, 0.03, 0.07, 0.11, 0.15, 0.19, 0.23, 0.26,  in (a) to (h), respectively. The central spot that stems from Rydberg states was covered in black for illustration purposes. The shown PMDs are projections from the range $|p_z|<0.06$ a.u. onto the polarization plane. These PMDs for ellipticities of $\epsilon = 0.07$ and $\epsilon = 0.11$ reveal a sharp line structure that disappears for larger ellipticities.}
	\label{fig:ellScan3D}
\end{figure*}
We recorded 3D PMDs at various ellipticities at an intensity of $6 \cdot 10^{13}$ W/cm$^2$ to study the ellipticity-dependence of rescattering in mid-IR laser pulses.
When the ellipticity is varied, the electron dynamics changes mainly in the polarization plane, i.e. the $(p_x, p_y)$-plane. To have access to the polarization plane, we record full 3D PMDs from strong-field ionization. The full 3D PMD is obtained by applying a tomographic reconstruction algorithm to the projected PMDs measured with velocity map imaging \cite{Wollenhaupt_2009, Smeenk_2009momentum, Dimitrovski_2014}.
The orientation of the 3D PMD with respect to the detector plane is linked to the orientation  of the polarization ellipse, allowing us to rotate the PMD by rotating the polarization ellipse. The beam propagation axis is parallel to the detector plane.
The polarization is rotated in steps of two degrees and a photoelectron image is recorded for each angular step.
Subsequently, for each slice along the beam propagation direction, a filtered back-projection algorithm is applied for the tomographic reconstruction. 
  
An example of a measured 3D PMD is visualized as an isosurface in Fig. \ref{fig:Surface}. The isosurface exhibits two main lobes and a sharp ridge around $p_y = 0$. 

\begin{figure}
\includegraphics{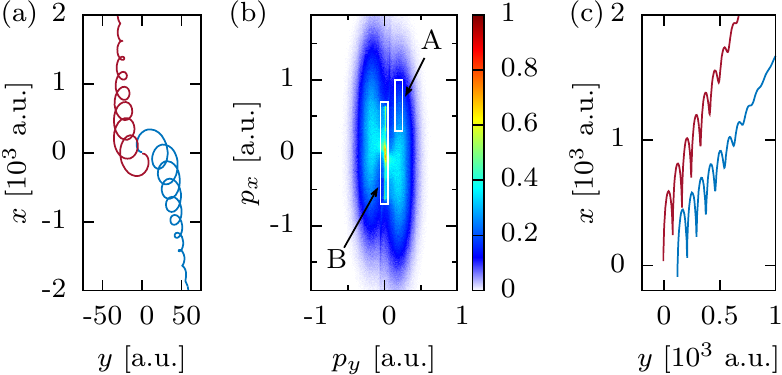}
\caption{Characteristic trajectories for different parts of the PMD. a) Characteristic trajectories from the central part of the PMD containing the focused photoelectrons (type B).  The trajectories revisit $x=0$ multiple times and have a point of intersection in the $(x,y)$-plane. c) The photoelectrons from the outer part (type A) of the PMD go directly to the detector without revisiting $x=0$ and do not have a point of intersection in the $(x,y)$-plane.} 
\label{fig:Traj}
\end{figure}

In the following we distinguish between two types of photoelectrons:

Type A photoelectrons: The ellipticity dependence of the PMDs in the polarization plane is shown in Fig. \ref{fig:ellScan3D}.
Cuts of the 3D PMDs through the polarization plane 
(i.e. $W(p_x, p_y) = \int_{\Omega_z} W(p_x,p_y,p_z) dp_z$) are shown in Fig. \ref{fig:ellScan3D} for ellipticities of 0.03, 0.07, 0.11, 0.15, 0.19, 0.23, 0.26. For the cuts we integrated over a range of $\Omega_z = |p_z|<0.06$ a.u.. The cuts show that, with increasing ellipticity, the cigar-shaped PMD evolves into a torus-like shape that is characteristic for strong field experiments with elliptically polarized pulses, such as typically observed with attoclock experiments \cite{Eckle_2008a, Eckle_2008b}. 
The appearance of the two maxima on the short axis of the polarization ellipse can be explained by a simple man's model \cite{SimpleManModel}.
Throughout this article, we will refer to the electrons ending in these maxima as type A electrons, as indicated in Fig. \ref{fig:Traj}.
Within the framework of this model, the maxima are shifted by 90$^\circ$ with respect to the phase at which the maximum of the electric field occurs. 
Deviations from 90$^\circ$ that are expected from the simple man's model are due to the Coulomb-interaction of the electron with the ion core \cite{Bashkansky_1988,Eckle_2008b, Pfeiffer_2012}, ionization delay times \cite{Eckle_2008a, Eckle_2008b, Landsman_2014o} and multi-electron effects like the induced dipole moment due to the ion’s polarizability \cite{Pfeiffer_2012}.

Type B photoelectrons:  In the evolution of the PMDs, one can observe for small elliptictities, in particular for $\epsilon = 0.07$ and $\epsilon = 0.11$, the appearance of a sharp, thin line-shaped ridge structure around $p_y = 0$. 
To the best of our knowledge, no such sharp separated structure has been observed in near-IR-experiments conducted at wavelengths around 800 nm. For the rest of this article, we will refer to these electrons as type B electrons.  

\begin{figure}
\includegraphics{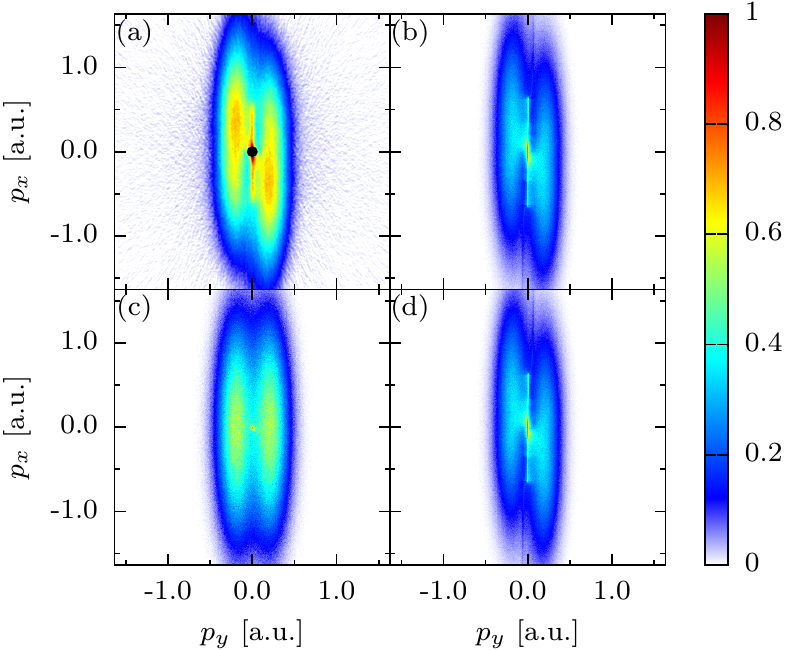}
\caption{Polarization plane PMDs projected from the range $|p_z|<0.06$ a.u.: Comparison of measurement (a) and CTMC calculations (b) from strong field ionization of xenon at an intensity of $6 \cdot 10^{13}$ W/cm$^2$ and an ellipticity of $\epsilon = 0.07$. For both experiment and simulation a sharp line appears around $p_y = 0$. (c) CTMC simulation for which the Coulomb potential was neglected, but the magnetic field component of the laser field was included. The sharp line around $p_y =0$ disappears. (d) CTMC simulation for which the Coulomb potential was included, but the magnetic field component was neglected.} 
\label{fig:Compar}
\end{figure}

To understand the nature of type B electrons, we compare them with CTMC simulations  using the two-step model of strong-field ionization. The initial conditions for the photoelectrons (ionization times, positions and momenta) are obtained from tunnel ionization theory in parabolic coordinates \cite{PPT,ADK,Delone_1991, TunnelIonization, Pfeiffer_2012}, while the trajectories of the electrons are obtained by solving Newton’s classical equations of motion in the electromagnetic field of the laser pulse and the Coulomb potential of the parent ion. Subsequently the trajectories are binned in momentum space.
We first compare the outcome of the simulations with our experiments in the polarization plane. The results are shown in Fig. \ref{fig:Compar}. The semiclassical simulations were able to reproduce the appearance of the type B photoelectron signal. Our semiclassical approach does not take into account
the quantum interference of the electron trajectories in the continuum, and thus we can conclude that the appearance of this structure is due to momentum space focusing of photoelectrons and is not created by a pure interference effect.

In order to understand if the creation of the ridge of type B electrons was induced by magnetic field effects, that appear at our beam parameters \cite{Ludwig_2014}, we performed the CTMC-calcuations with and without the inclusion of the magnetic field component (Fig. \ref{fig:Compar} (b) and (d)). The CTMC calculations reproduce the sharp structure of type B-electrons in the $(p_x, p_y)-$plane equally well. Thus, we conclude that for the appearance of type-B electrons in the polarization plane the magnetic field is not essential.   
Furthermore, to confirm that the structure was created under the influence of the Coulomb-potential of the ion, we performed CTMC-simulations with and without the Coulomb-potential included. It is obvious from the comparison of Figs. \ref{fig:Compar} (c) (Coulomb-potential not included in simulation) and (d) (Coulomb-potential  included) that the inclusion of the Coulomb potential is required to reproduce the experimental data. 

Type A and type B electrons have different characteristic position-space trajectories as shown in Fig. \ref{fig:Traj}. Type A electrons travel directly to the detector without revisiting $x=0$ and do not have a point of intersection in the $(x,y)$ plane. However, type B electron trajectories have a point of intersection in the $(x,y)$ plane and furthermore, they cross $x = 0$ multiple times. 

\section{Coulomb focusing at elliptical polarization}
\subsection{Creation of the sharp, thin line-shaped ridge structure in the 3D PMD}
\label{sec:AnaOne}

\begin{figure}
\includegraphics{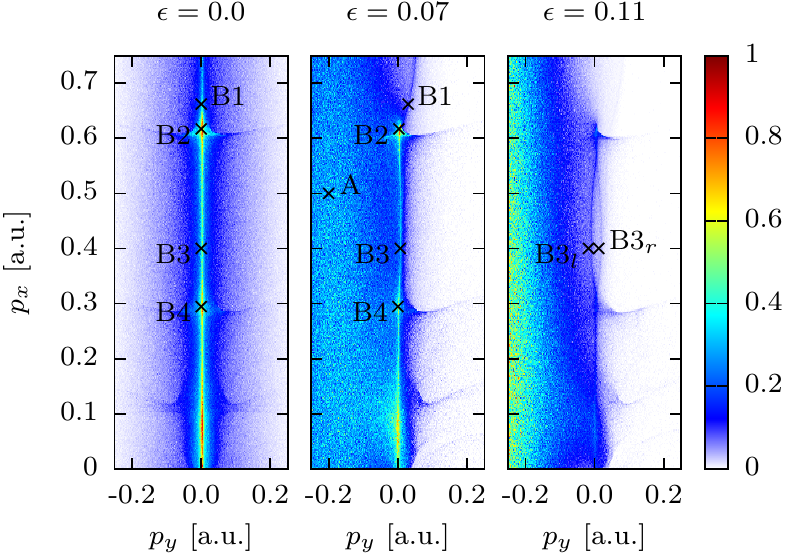}
\caption{Comparison of central parts of PMDs obtained by CTMC for different ellipticities: $\epsilon = 0$, $\epsilon = 0.07$ and $\epsilon = 0.11$, respectively. 
Characteristic points of the PMD depending on the specific longitudinal momenta, correspond in the linear polarization case,  (B2) to the trajectory with  two rescatterings, (B3)  with three,  (B4)  with four rescatterings, and (B1) with a single rescattering. (B2) and (B4) correspond to the slow recollision condition ($p_{xr}=0$).  (A) corresponds to the center of the main lobe and is originating from recollision-free trajectories.
Horizontal caustics are visible around, e.g., the characteristic points B2 and B4. Furthermore, a vertical caustic due to Coulomb focusing is visible as the line through the points $Bn$.    
For $\epsilon = 0.11$, the vertical caustic at point B3 is split.} 
\label{fig:01-comparison_of_PESs}
\end{figure}

To understand the nature of this ridge structure in more detail, we analyze the 3D PMD in the polarization plane (i.e. in the $(p_x, p_y)$-plane). We begin the analysis in the simpler case of a linearly polarized laser field and transfer the results to the case of elliptical polarization with  $\epsilon>0$. Fig. \ref{fig:01-comparison_of_PESs} shows the results of CTMC simulations in dipole approximation for three different ellipticities: $\epsilon = 0$, $0.07$ and $0.11$. To focus on the mechanism of the creation of the ridge, we consider only electrons starting in the central half-cycle of the laser field to suppress the influence of ionization from multiple half cycles. Furthermore, we neglect the magnetic field effects to disentangle the creation of the ridge in the $(p_x, p_y)$-plane from additional effects in beam propagation $p_z$-direction. 

As we saw in section \ref{sec:PMD3D}, the ridge already appears within the dipole approximation because the CTMC simulated PMD in the polarization plane did not depend on the magnetic field.
In the case of linear polarization, the features of the PMD are   understood in the terms of laser-driven classical trajectories recolliding with the parent ion. Due to the nature of the Coulomb interaction, a bunching of electrons occurs and is imprinted on the PMD in the form of caustics (Fig.~\ref{fig:01-comparison_of_PESs}). There are two kinds of caustics: horizontal and vertical. Each horizontal caustic line in the PMD corresponds to a certain class of rescattered trajectories. When the longitudinal momentum $p_x$ of the electron at the recollision is vanishing, i.e. the slow recollision condition  is fullfilled,   longitudinal bunching of electrons occurs \cite{Kastner_2012}. This condition depends on the ionization phase (i.e. the phase of the laser electric field when the electron appears in the continuum).

In Fig. \ref{fig:Traj}, we indicate several characteristic points exhibiting qualitatively different rescattering behavior. Point A is an example of a type A electron, while points $Bn$ indicate type B electrons with n rescattering events. For these cases the sharp peak in the distribution around $p_y=0$, can be seen due to Coulomb focusing. This peak is pronounced for small ellipticities, but starts to disappear for $\epsilon \gtrsim 0.1$. 

\begin{figure}
	\centering
	\includegraphics{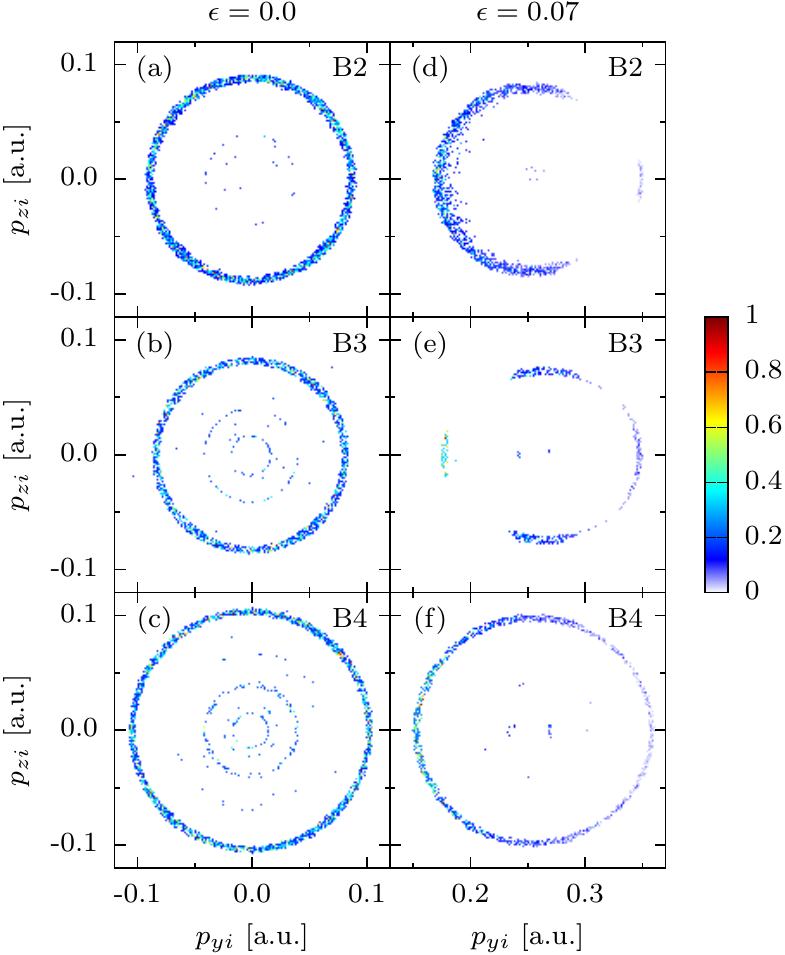}
	\caption{The initial momentum space distribution for trajectories ending in the momentum bins of $0.01\times0.01\times0.01$ dimensions at the asymptotic values of longitudinal momentum $p_x$: (a)  $0.62$, (b) $0.44$ and (c) $0.3$ $a.u.$,  corresponding to the points B2, B3, B4 of Fig.~\ref{fig:01-comparison_of_PESs}, respectively.   The laser  field is linearly polarized in (left column), and  elliptically polarized, with $\epsilon = 0.07$, in (right column).  Notice the positive offset in the $p_{yi}$ of the structures in the right column due to the ellipticity.}
	\label{fig:02-initial_phase_space-e_0_and_0_07_P1_P2_P3}	
\end{figure}

In order to understand this vertical ridge further, we analyze the set of initial transverse momentum distributions (i.e. at the tunnel exit) corresponding to the final transverse momentum values given by points $Bn$. The resulting distributions are shown in Fig. \ref{fig:02-initial_phase_space-e_0_and_0_07_P1_P2_P3}. We  analyze the initial transverse momentum distribution at the tunnel exit with momentum bins of $0.01\times0.01\times0.01$ a.u. placed at characteristic points of the PMD. We compare the initial PMDs of corresponding points for linear and elliptical polarization (Fig.~\ref{fig:02-initial_phase_space-e_0_and_0_07_P1_P2_P3}). 

The ridge originates from a contraction in momentum transverse to the major polarization axis. In a linearly polarized laser field, the electrons contributing to the ridge are ionized with a nonvanishing transverse momentum at the tunnel exit, and appear after propagation with a vanishing transverse momentum. Their initial distribution is a ring in the ($p_y,p_z$)-initial momentum distribution, see Fig.~\ref{fig:02-initial_phase_space-e_0_and_0_07_P1_P2_P3}, left column. 
in the case of a linearly polarized laser field, the electrons which are initially (i.e. at the tunnel exit) distributed inside this ring obtain a large momentum transfer during recollisions and end up outside of the chosen final momentum bin near the vanishing transverse momentum.

In the case of the elliptically polarized laser field, there are two modifications to this picture. For small ellipticities (quantified in Eq.~(\ref{small-ellipt}) below) the rescattering and Coulomb focusing, similar to the case of linear polarization, takes place for electrons which initially are distributed in a {\it shifted} ring of initial momenta in the  ($p_{yi},p_{zi}$)-plane.  The size of the shift is discussed further in section \ref{sec:AnaTwo}). The radius of the ring of the initial momentum space distribution is an indicator for Coulomb focusing. It is nearly the same for linear and elliptical polarization, i.e., Coulomb focusing for these trajectories is qualitatively the same. 
The points B2 and B4 in Fig.~\ref{fig:01-comparison_of_PESs} corresponding to the slow recollision condition do not change their position in the PMD when changing ellipticity, which is due to the similarity of the underlying trajectories.

Next, we analyze these trajectories analytically to show that the Coulomb focusing dynamics is  similar for linear and elliptical polarization up to a certain value of ellipticity (especially in the case of slow recollisions). Furthermore, we show that the recolliding electrons, which create the ridge, end up around $p_{yf} \approx 0$.  

The underlying electron trajectories are obtained from the solution of the electron equations of motion in an elliptically polarized laser field assuming that the Coulomb field effect is a perturbation, which affects the electron trajectory near the tunnel exit and at recollisions. The electric field component of the laser field is
\begin{eqnarray}
E_x &=& E_0\cos \eta
\label{eq:laserField1}\\
E_y &=&\epsilon E_0\sin \eta.
\label{eq:laserField2}
\end{eqnarray}
with the phase $\eta $, the ellipticity $0\leq\epsilon\leq 1$, the field amplitude $E_0=\sqrt{I}/\sqrt{1+\epsilon^2}$, and the intensity $I$. The envelope of the pulse is neglected.  For  the electron  dynamics in the laser polarization plane after the ionization, taking into account initial Coulomb momentum transfer at the tunnel exit, we have:
\begin{eqnarray}
p_x &=& -\frac{E_0}{\omega}\left(\sin \eta-\sin\eta_i\right)-\delta p_{xi}^C\label{px}\\
p_y &=&  \epsilon\frac{E_0}{\omega}\left(\cos \eta-\cos\eta_i\right)+p_{yi}-\delta p_{yi}^C,
\label{elliptt-mom-y}
\end{eqnarray}
where $\eta_i$ is the ionization phase, $p_{yi}$ is the initial transverse electron momentum, and $\delta p_{xi}^C$ and $\delta p_{yi}^C$ are initial Coulomb momentum transfer. 
The electron rescattering and Coulomb focusing in an elliptically polarized laser field will be similar to the case of linear polarization, when the Coulomb momentum transfer during recollision is the same in both cases: $\delta p_{yr}^{C(\epsilon)}=\delta p_{yr}^{C(0)}$. 
For the latter it is necessary to have the same impact parameter of the recollision, i.e. the same recollision $y$-coordinate. This condition is fulfilled when the initial momentum at the tunnel exit is shifted with respect to the linear polarization case by a value to compensate the momentum imparted to the electron by the $y$-component of the laser field (as well as a small difference of the initial Coulomb momentum transfers). 
The value of this shift at the slow recollision condition is: 
\begin{eqnarray}
p_{yi}^{(\epsilon)}  -p_{yi}^{(0)}=\epsilon\frac{E_0}{\omega }\cos\eta_i +\delta p_{yi}^{C(\epsilon)}-\delta p_{yi}^{C(0)}+\frac{\epsilon\delta p_{xi}^{C(\epsilon)}}{\eta_r-\eta_i} ,
 \label{pyi}
\end{eqnarray}
see the derivation in the appendix \ref{sec:app}. Here, the superscripts $(\epsilon)$ and $(0)$ refer to the cases of elliptical and linear polarization, respectively. The final momentum of the type B electron is found by combining Eqs.~(\ref{elliptt-mom-y}) and (\ref{pyi}):
\begin{eqnarray}
 p_{yf}^{(\epsilon)} =-\epsilon\frac{E_0}{\omega }\cos\eta_i+p_{yi}^{(\epsilon)}-\delta p_{yi}^{C(\epsilon)} -\delta p_{yr}^{C(\epsilon)} = \frac{\epsilon\delta p_{xi}^{C(\epsilon)}}{\eta_r-\eta_i },
\label{pyf}
\end{eqnarray}
where we have taken into account that  $\delta p_{yr}^{C(\epsilon)}=\delta p_{yr}^{C(0)}$ and $p_{yi}^{(0)}=\delta p_{yi}^{C(0)}+\delta p_{yr}^{C(0)}$, i.e. that in the case of linear polarization the main ridge due to Coulomb focusing is at $p_y\approx 0$.  The final momentum of the type B electron  has practically a vanishing value. In fact, $\delta p_{xi}^{C(\epsilon)}\approx \pi E(\eta_i)/(2I_p)^{3/2}$ \cite{Shvetsov-Shilovski_2009}, and $\eta_r-\eta_i\sim 3\pi$ at the first slow recollision, and $\epsilon\delta p_{xi}^{C(\epsilon)}/(\eta_r-\eta_i)\sim 10^{-3}$, at $\epsilon\sim 0.1$, where $\eta_r$ is the phase of recollision. All numerical estimations in this section are for $\omega=0.013$ ($\lambda=3400$ nm), $E_0=0.04$ (intensity $5.8\times 10^{13}$ W/cm$^2$).
Thus, neglecting initial Coulomb momentum transfer for simplicity, we can conclude that in the case of elliptical polarization, the electrons with slow recollision condition are initially distributed in momentum  space ($p_{yi},p_{xi}$) on the ring centered at 
\begin{eqnarray}
p_{yi}^{(\epsilon)}\approx\epsilon\frac{E_0}{\omega }\cos\eta_i,\,\,\,\,
p_{zi}^{(\epsilon)}=0, \label{mshift}
\end{eqnarray}
with a radius $\delta p_{yr}^{C(\epsilon)}$
and finally will end up at the ridge around $p_y \approx 0$. 
This is illustrated in  Fig.~\ref{fig:02-initial_phase_space-e_0_and_0_07_P1_P2_P3}, right column, where the top (B2) and bottom (B4) panels correspond to the slow recollision condition. 
The radii of the rings in B2 and B4, which indicate the magnitude of Coulomb focusing, do not change significantly due to the change in ellipticity. Thus, we can conclude that Coulomb focusing dynamics is very similar in both cases.

\begin{figure}
	\centering
	\includegraphics{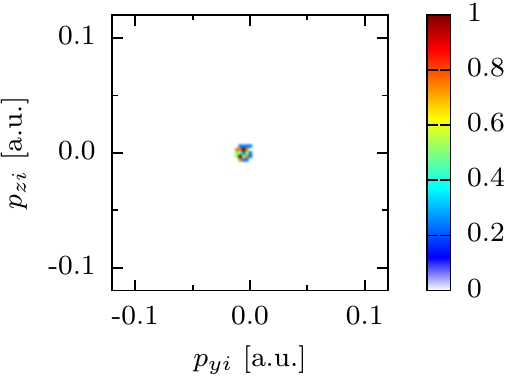}
		\caption{The initial momentum space distribution of direct electrons without Coulomb focusing which are ending in the center of the main lobe,  point A in Fig.~\ref{fig:01-comparison_of_PESs}. }
	\label{sidelobe}
\end{figure}

At non-negligible ellipticities the electrons around vanishing initial transverse momentum experience no recollision, no momentum change due to the Coulomb field besides the initial Coulomb momentum transfer when the electron leaves the tunnel exit. They contribute to the lobes of the final $(p_x,p_y)$-distribution, with the final momentum $\textbf{p}_f\approx -\textbf{A}(t_i)+\delta \textbf{p}_{i}^C$, with the  initial Coulomb momentum transfer $\delta \textbf{p}_{i}^C$.  The latter is mostly along the electric field for the experimental parameters (the transverse component of initial Coulomb momentum transfer is smaller with respect to the longitudinal component by an order of magnitude), i.e., perpendicular to the vector potential, and induces a distortion of the PMD ellipse with respect to the case of the simple man's model [$\textbf{p}_f\approx -\textbf{A}(t_i)$].
The initial momentum space corresponding to the center of the main lobe A is shown in Fig.~\ref{sidelobe} and indicates the absence of Coulomb focusing (the initial and the final phase space are the same). 

Next we want to estimate up to which ellipticities the central sharp ridge of Coulomb focusing electrons still appears in the PMD.   
The discussion above  is valid if the required initial transverse momentum of type B electrons according to Eq.~(\ref{pyi}), $p_{yi}^{(\epsilon)}\sim \epsilon E_0/\omega+\delta p_{yr}^{C(\epsilon)}$, is within the momentum width tunnelled electron wave packet. The latter reads 
\begin{eqnarray}
\epsilon\frac{E_0}{\omega }\lesssim \Delta_\bot, 
\label{epsilon-central}
\end{eqnarray}
where $\Delta_\bot=  \sqrt{E_0}/(2I_p)^{1/4}$  is the PMD width ($2\sigma$ of the Gaussian distribution) at the tunnel exit according to tunnel-ionization theory\cite{PPT,ADK} 
With Eq.~(\ref{epsilon-central}) we conclude that the ridge in the PMD can exist up to ellipticities
\begin{eqnarray}
\epsilon\lesssim \frac{\omega }{\sqrt{E_0}(2I_p)^{1/4}}\approx 0.07.  
  \label{small-ellipt}
  \end{eqnarray}

There are also modifications of the rings in Fig.~\ref{fig:02-initial_phase_space-e_0_and_0_07_P1_P2_P3}, especially for B3, which we discuss later in the next section based on trajectory analysis of these characteristic points.
Note that the inner rings in the linear case in Fig.~\ref{fig:02-initial_phase_space-e_0_and_0_07_P1_P2_P3} are caused by trajectories with multiple significant rescattering events. However, such trajectories are strongly suppressed for increasing ellipticity. 

Finally, let us estimate at which ellipticity the side lobes will be separated from the sharp ridge in the PMD. The side lobes appear when the ellipticity is large enough such that the Coulomb momentum transfer at recollision, for the electron with initial $p_{yi}=0$, is negligible with respect to the final momentum.  The final electron momentum  can be estimated as
\begin{eqnarray}
p_{yf} \approx  \epsilon \frac{E_0}{\omega},
\end{eqnarray}
and the Coulomb momentum transfer as
\begin{eqnarray}
\delta p_{yr}^C\approx -\frac{y_r}{R_r^3} \delta t_r\approx \frac{\delta t_r}{y_r^2},
\end{eqnarray}
where $x_r$, $y_r$, $z_r$, $R_r=\sqrt{x_r^2+y_r^2+z_r^2}$ are the electron coordinates, and the distance from the core at the recollision point, respectively. $\delta t_r$ is the recollision duration:
$\delta t_r\approx \sqrt{2y_r/E_x(t_r)}$.
For the latter, we assumed that during the recollision time the electron travels a distance of the order of $R_r$ in x-direction , i.e., $\delta x\approx R_r\approx y_r\approx E_x(t_r)\delta t^2/2$ when we apply the condition for slow recollision. Estimating the  recollision coordinate as
$y_r\approx \epsilon  E_0/\omega^2$,
we have for the Coulomb momentum transfer
$\delta p_{yr}^C\approx \sqrt{2}\omega^3/(\epsilon^{3/2}E_0^2)$.
Thus, Coulomb focusing at recollision will be negligible, $\delta p_{yr}^C\ll p_{yf}$, if the ellipticity is relatively large
\begin{eqnarray}
 \epsilon \gtrsim \left(\frac{\sqrt{2}\omega^4}{E_0^3}\right)^{2/5}\approx 0.05.
 \label{eq:cond_on_epsilon_for_side-lobes}
\end{eqnarray}
 At  ellipticities larger than this value, the lobes in the $(p_x, p_y)$-plane begin to appear.

\begin{figure}
	\centering
	\includegraphics{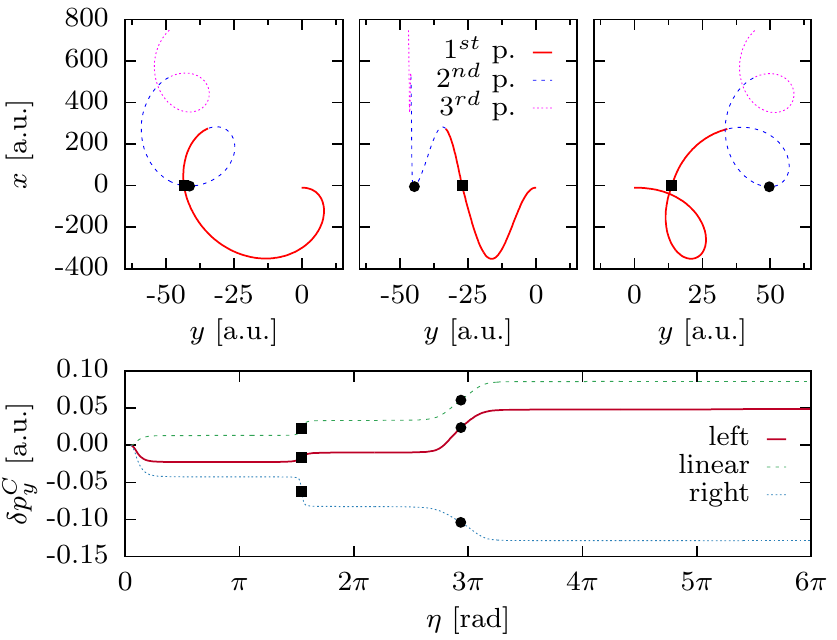}
	\includegraphics{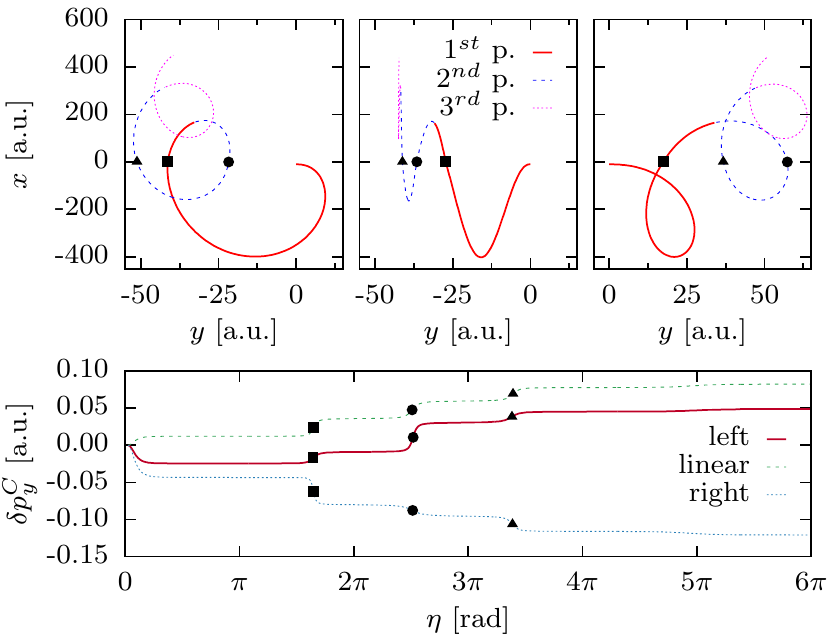}
	\includegraphics{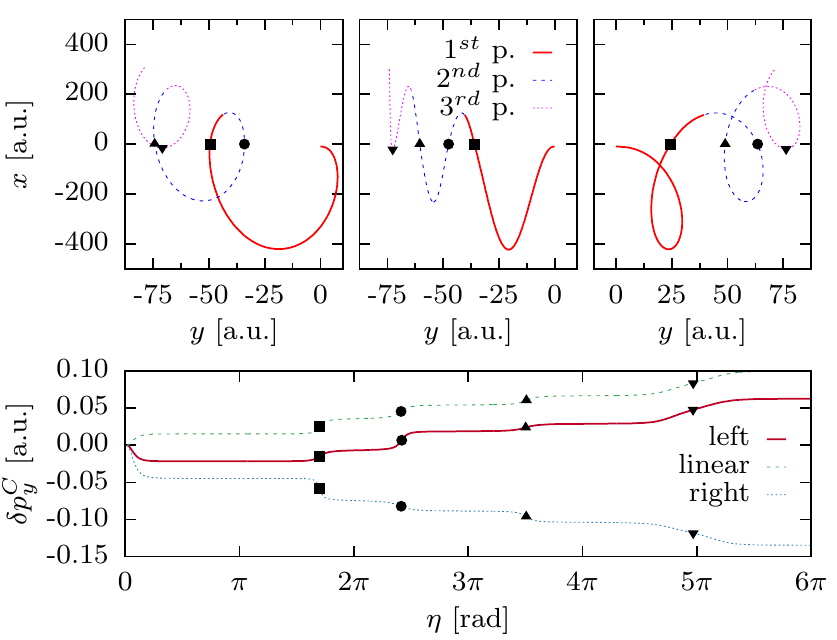}
		\caption{Typical photoelectron trajectories in a laser field with $\epsilon = 0.07$, and Coulomb momentum transfers initially and during subsequent recollisions. The trajectories originate on the left (left panels) and right part (right panels) of the initial transverse momentum distribution ring with $p_{zi} = 0$ and end up at the same point B2 (upper group), B3 (middle group), and B4 (bottom group).  The middle panels of trajectories show the case of linear polarization.}
	\label{fig:03-trajectories-e_0_07-P1}
\end{figure}

\subsection{The structure of the sharp, thin line-shaped ridge in the 3D PMD}
\label{sec:AnaTwo}

To investigate the structure of the sharp ridge, (Fig.~\ref{fig:Surface}), we analyse in this section the corresponding trajectories. The structure of the ridge is very similar to the linear polarization case at the slow recollision condition. The structure slightly deviates from the linear polarization case outside of the slow recollision condition.

In Fig.~\ref{fig:03-trajectories-e_0_07-P1}, we analyze typical trajectories corresponding to each of the points B2, B3, B4, and originating from the left $(-)$ and right $(+)$ parts of the ring of initial momentum distribution (Fig.~\ref{fig:02-initial_phase_space-e_0_and_0_07_P1_P2_P3}). We choose left and right points which have the same final vanishing momentum,
\begin{eqnarray}
p_{yf}^-=p_{yf}^+.\label{==}
\end{eqnarray}

The difference of the initial momenta of these points determines the diameter of the ring of the initial momentum distribution  of Fig.~\ref{fig:02-initial_phase_space-e_0_and_0_07_P1_P2_P3}, which depends on Coulomb momentum
transfer during recollision, and is an indicator of Coulomb focusing: 
\begin{eqnarray}
D\equiv p_{yi}^+-p_{yi}^-=\left(\delta p_{yr}^{C+} + \delta p_{yr}^{C-}\right)\left(1+\frac{2E(\eta_i)}{(2I_p)^2} \right),\label{DDD}
\end{eqnarray}
see the derivations for this section in Appendix \ref{sec:app2}. The center of the ring is given by 
\begin{eqnarray}
\frac{p_{yi}^++p_{yi}^- }{2} = \left(\epsilon\frac{E_0}{\omega}\cos\eta_i+\frac{\delta p_{yr}^{C+}}{2}-\frac{\delta p_{yr}^{C-}}{2} \right) \left(1+\frac{2E(\eta_i)}{(2I_p)^2} \right), 
\end{eqnarray}
which includes initial Coulomb momentum transfer and Coulomb momentum
transfer during recollision correction into Eq.~(\ref{mshift}), and indicates that in the case of elliptical polarization the ring of the initial momentum distribution for the sharp ridge electrons is shifted due to the drift momentum along the minor axis of polarization, and by initial Coulomb momentum transfer in that direction as $\delta p_{yr}^{C+}\approx \delta p_{yr}^{C-}$:  
\begin{eqnarray}
p_{yi}^{(\epsilon)}
 \approx \epsilon\frac{E_0}{\omega }\cos\eta_i\left[1 +\frac{2 E_0\cos\eta_i}{(2I_p)^{2}}\right], 
\end{eqnarray}
where $p_{yi}^{(\epsilon)}$ describes the shift of the initial momentum along the $p_y$-axis at an ellipticity $\epsilon$ compared to the linear case. 
For an ellipticity of $\epsilon = 0.07$, we have $p_{yi}^{(\epsilon)}
 \approx 0.23$.

\begin{figure}
	\includegraphics{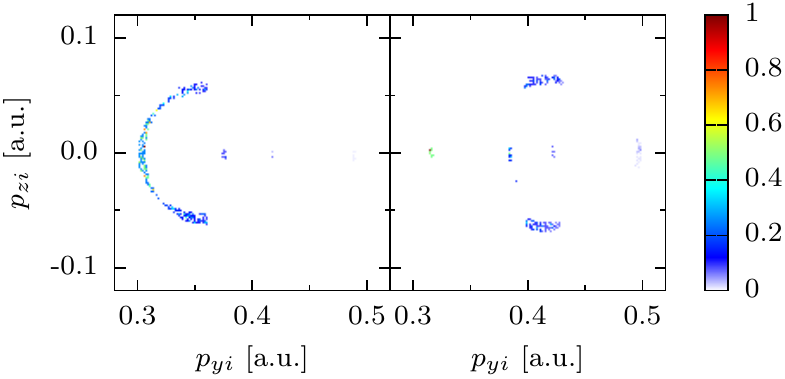}
	\caption{The initial momentum distributions for $\epsilon = 0.11$ leading to the two distinct ridges [B3(l) and B3(r)] taken at the same value of longitudinal momenta. The pronounced left branch consists of electrons following linear-like trajectories, whereas the faint right branch comes from trajectories strongly influenced by the ellipticity. }
	\label{fig:07-initial_phase_space-e_0_11_P2}
\end{figure}

The diameter ($D$) of the ring in the initial momentum distribution, the indicator of Coulomb focusing, depends on Coulomb momentum
transfer during recollision, but not on initial Coulomb momentum transfer $\delta p_{yi}^{C\pm }$ (see appendix \ref{sec:app2}, Eq.~(\ref{DDD})). For all three points B2, B3, B4, the Coulomb momentum
transfer during recollision is approximately the same as in the case of linear polarization, cf. Fig. \ref{fig:03-trajectories-e_0_07-P1}. Therefore, the radius of the ring in the initial transverse momentum distribution and the ridge in the final PMD are also approximately the same. 

We next analyze the modifications of the ridge with the variation of ellipticity. For the points B2 and B4 and linear polarization the main Coulomb momentum
transfer during recollision takes place at the slow recollision,  $p_{xr}=0$. This corresponds to the second recollision for B2, and the fourth one near B4, although there are multiple recollision points, $x_r=0$, two at B2, and four at B4 (see Fig. \ref{fig:03-trajectories-e_0_07-P1}). In the elliptical polarization case, the Coulomb focusing for the trajectory coming from the left part of the ring resembles the linear case, since the slow recollision has the same impact parameter and the same Coulomb momentum
transfer during recollision. Of the multiple recollisions that occur for these points, the slow one has the dominant effect on the final momentum. Therefore, the Coulomb focusing for the left half-ring in Fig.~\ref{fig:02-initial_phase_space-e_0_and_0_07_P1_P2_P3} is the same for both linear and elliptical polarizations, thereby creating the central ridge in the PMD in both cases.

In contrast, the trajectory from right part of the ring at B2 in the elliptical polarization case differs from the linear one, see Fig.~\ref{fig:03-trajectories-e_0_07-P1}, upper panel. The first rescattering for this trajectory, additional to the slow recollision, takes place with shorter impact parameter than in the linear case due to the oscillating part of the $y$-coordinate, yielding to the increase of the total Coulomb momentum transfer. This explains the larger radius of the right half-cycle of the ring structure at B2 in Fig.~\ref{fig:02-initial_phase_space-e_0_and_0_07_P1_P2_P3}. The right-type trajectories are more sensitive to the initial conditions and, consequently, the ring width is significantly smaller. Moreover, the ionization probability is smaller for the right part of the ring because of the lager initial momenta. These explain why the central ridge becomes less pronounced with increasing ellipticity. 

However, the right trajectories for B4, and trajectories ending up on the central ridge at $|p_x|$ smaller than for B4,  start to resemble the linear case again. 
This is because for the ionization phases that are relevant to the latter trajectories,  the oscillating part of the $y$-coordinate does not perturb the recollision coordinates significantly. The condition that the recollision coordinate is only weakly perturbed by the ellipticity, i.e. $|y_r^{(\epsilon)}-y_r^{(0)}|\ll y_r^{(0)}$,
leads to an estimate of the threshold ionization phase (see appendix \ref{sec:app3}):
\begin{eqnarray}
\eta_i\ll \pi \left(\frac{\delta p_{y}^{C}}{(\epsilon E_0/\omega)} \right)^2\approx 0.68,
\end{eqnarray}
for $\delta p_{y}^{C} = 0.1$. The numerical value in the equation above is estimated for the parameters of Fig.~\ref{fig:01-comparison_of_PESs}, i.e., at $\eta_i\lesssim 0.068$, which assumes $p_{x}\lesssim E_0\eta_i/\omega \approx 0.2 $, the recollisions and Coulomb focusing in the elliptical polarization case will be similar to the linear one. This estimation fits to the CTMC calculation in Fig.~\ref{fig:01-comparison_of_PESs}.

The ring of the initial momenta of point B3 is deformed in a stronger way then the one for B2. The deformation is due to the fact that both left and right trajectories are perturbed with respect to the linear polarization case because of the quiver motion in the transversal $y$-direction. The perturbed trajectories are more sensitive to the initial conditions, which results in a variable width of the ring in the initial momentum distribution. Moreover, the recollision coordinates for both left and right trajectories are different. Therefore, the Coulomb momentum
transfer during recollision for the left and right trajectories are not symmetric which leads to the bend of the central ridge.  Furthermore, at larger ellipticities, e.g. $\epsilon=0.11$, the central ridge at B3 is split, when the left- and right-side trajectories yield to different ridges, see Fig.~\ref{fig:07-initial_phase_space-e_0_11_P2}. However, this splitting is not visible in the experimental data due to focal volume averaging, CEP averaging and the laser pulse shape.

\section{Non-dipole effects}
\label{sec:noDipole}

The PMDs in Ref. \cite{Ludwig_2014} recorded at mid-IR wavelengths showed a strong influence of the Coulomb-potential in combination with the magnetic field component of the laser field due to rescattering.

\begin{figure*}
     \includegraphics{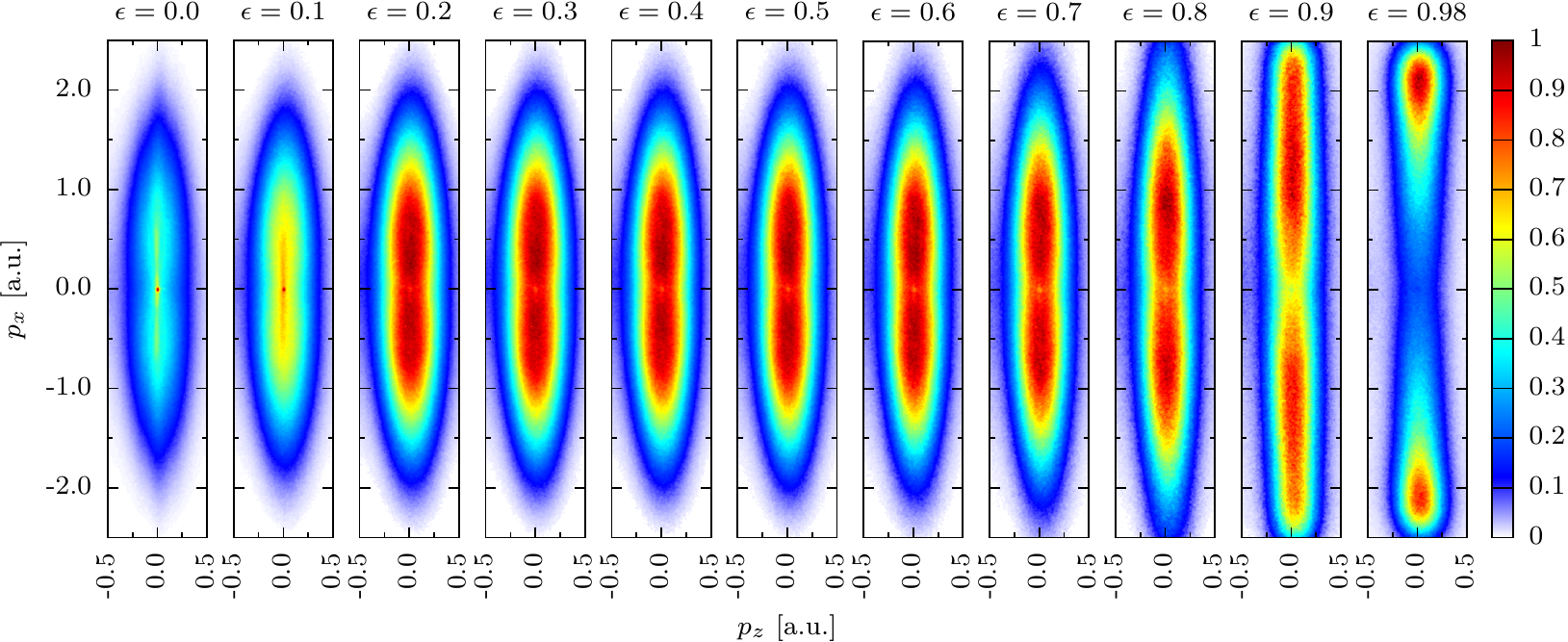}
		  \caption{Normalized measured 2D PMDs from xenon recorded for various ellipticities at a peak intensitiy of $6 \cdot 10^{13}$ W/cm$^2$.}
        \label{fig:ellScan1}
  \end{figure*}
  
\begin{figure}
\includegraphics{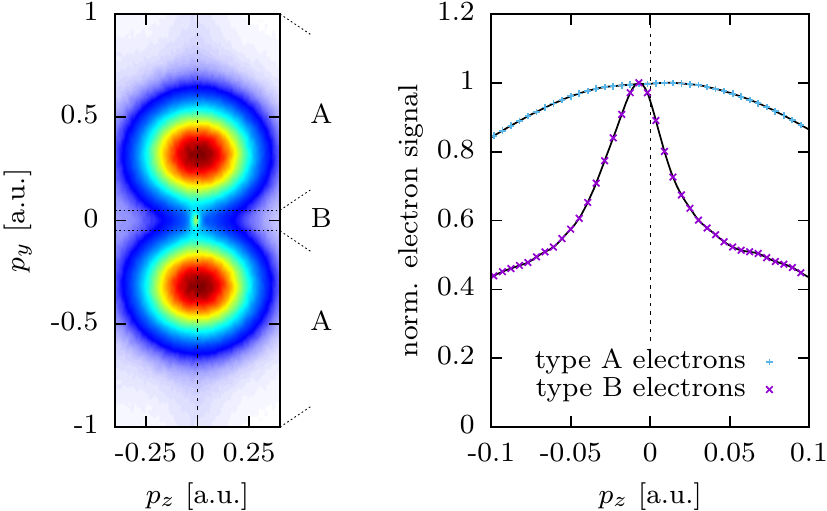}
\caption{Projection of 3D PMD for $p_x<0.05$ a.u. recorded at a peak intensity of $6 \cdot 10^{13}$ W/cm$^2$ and an ellipticity of $\epsilon =0.11$ (left). The sharp line structure is clearly visible around $p_y = 0$. The ranges of $p_y$ for the projection in are marked with dashed lines. The central spot stemming from ionization of Rydberg atoms was removed prior to the projections (see main text for details).
Projections of the photoelectrons onto the beam propagation axis for two different regions of $p_y$ (right). The magenta crosses show the photoelectron distribution  for $|p_y|<0.05$ a.u. The blue crosses show the distribution  for $|p_y|>0.05$ . The peak of the inner electrons is shifted opposite to the beam propagation direction whereas the outer electrons are shifted in beam propagation direction. The black lines serve as a guide for the eye.} 
\label{fig:Shift3D}
\end{figure}

In order to minimize the influence of rescattering processes, we studied the projection of the PMD onto the beam propagation axis for the case of close-to-circular polarization for helium and xenon. We extract the zero momentum from reference images recorded with linear polarization as described in section \ref{sec:exp}. 
The peaks of these projections were extracted via a Gaussian fit to the projection of the PMD onto the $p_z$-axis. The results are shown together with a model from Ref. \cite{Smeenk_2011} for circular polarization in Fig. \ref{fig:ellShift}. 
The results for circular polarization are consistent with the results from Ref. \cite{Smeenk_2011} within the error bars. Theoretical studies predicted an additional offset of the order of $\frac{I_p}{3c}$ at the tunnel exit \cite{Klaiber_2013c} as well as in the final momentum distributions \cite{Chelkowski_2014,Cricchio_2015,Chelkowski_2015}.
However, the experiment cannot resolve this additional offset. 
In addition we show for comparison the data for linear polarization from Ref. \cite{Ludwig_2014}. The offsets for the case of linear polarization are shifted in contrast to the data for circular polarization opposite to the beam propagation direction.

The results with linear and circular polarization (Fig. \ref{fig:ellShift} (a)) trigger the question what happens when the ellipticity is being changed? 
Since our previous results \cite{Ludwig_2014} indicate that the non-dipole effects are independent of the target gas within the accuracy of our measurement. We therefore perform our studies on xenon.
The ellipticity  was varied in steps from linear to close-to-circular ($\epsilon=0.97$). We ionized xenon atoms at an intensity of $6 \cdot 10^{13}$\,W/cm$^2$ with 50-fs pulses at a center wavelength of 3.4 $\mu$m. For each ellipticity step, projected momentum images in the ($p_x$, $p_z$) plane were recorded. We would like to point out that during the measurement we kept the intensity constant, not the electric field. That allows us to keep the total momentum transfer per cycle onto a free electron from the field independent of the ellipticity. 
Fig. \ref{fig:ellScan1} shows how the projected PMD is evolving from the typical cigar-like shape with the major dimension in $p_x$-direction towards a structure with maxima shifted to high values of $p_x$. Furthermore,  the spot in the center stemming from Rydberg-atoms ionized by the spectrometer field disappears with increasing ellipticity because the selection rules do not allow the excitation from the ground state into Rydberg states with circularly polarized light. 
Thus, reference measurements with linear polarization were taken right before and after each ellipticity step. The peaks for reference zero were extracted in a similar fashion as the method applied in Ref. \cite{Ludwig_2014} and the one described in section \ref{sec:exp}. For each reference measurement, we recorded $N=10$ photoelectron momentum images. The uncertainties of the determination of zero momentum scale with $\frac{1}{\sqrt{N}}$ due to $N$ repetitive measurements that were weighted with the error for a Gaussian fit of the peak of each projected distribution.

The zero momentum reference was extracted from the momentum images with linear polarization from the peak of the projection of a thin slice of $|p_x| \le 0.05$ a.u. around $p_x = 0$ projected onto the beam propagation. The peak of the projection was identified via a polynomial fit to the central part (i.e. $\Delta p_y \approx 0.05$ a.u.). of the PMD $W(p_z) = \iint_\Omega W(p_x,p_y,p_z) dp_x dp_y$, with $\Omega$ being the integration momentum volume. The offset of that peak from $p_z = 0$ is shown in Fig. \ref{fig:ellShift}(b) as a function of the ellipticity. 
We observe an increase of the offset with increasing ellipticity from negative values (i.e. opposite to the beam propagation direction) to positive values (i.e. in beam propagation direction). The transition from negative to positive values occurs at an ellipticity of $\epsilon = 0.12$. Thus, we have observed a zero-crossing along the $p_z$-axis of the peak of the projected PMD as we scan the ellipticity.  

Next, we consider the role of Coulomb focusing in the electronic response to non-dipole effects. 
Thus  we study the different responses of type A and type B electrons. Since we are able to isolate the electrons undergoing Coulomb-focusing in 3D momentum-space, we can study their response separately from the mostly unfocused electrons for $\epsilon\approx 0.1$.
We select type B – electrons by choosing a narrow momentum range of 0.038 a.u. in $p_y-$direction from the 3D PMD recorded at an ellipticity of $\epsilon = 0.11$. The electron signal that lies within this range and the electron signal from outside this range are separately projected onto the $p_z$-axis. 
The central spot stemming from ionization of Rydberg atoms was removed prior to the projections by removing the photoelectron signal with $|p|<0.03$ a.u..
The normalized projected electron signals are shown together in Fig. \ref{fig:Shift3D}. The position of the peaks from A and type B electron signals on the $p_z$-axis were identified. We observe that the type A electron signal peaks at a positive value of $p_z$ and the type B electron signal at a negative value of $p_z$. 

Increasing  ellipticity  supresses rescattering and thus, the PMD and its projection becomes dominated by electrons that interact only weakly with the Coulomb-potential of the parent ion. However, for ellipticities $\epsilon \lesssim 0.12$, the sharp ridge consisting of type B electrons starts to dominate the projection of the PMD onto the $p_z$-axis.  This creates a peak of electrons that significantly interact with the Coulomb-potential of the parent ion. That peak is shifted opposite to beam propagation direction.

\section{Conclusions}

In conclusion, we are able to separate the electrons that experienced strong Coulomb focusing in momentum space from the unfocused electrons in the case of elliptical polarization. 
With a detailed analysis, we identified a subspace in the momentum representation of the tunnel ionized electron wave packet, which in an elliptically polarized laser field shows recollision dynamics similar to  the linear polarization case. The electron trajectories originating in this initial PMD subspace have multiple revisits, including at least one significant rescattering event with the parent ion, which causes the large initial momentum subspace to be squeezed into a small final momentum space having near vanishing transverse momentum. 
This effect occurs due to Coulomb focusing, and leads to the creation of the sharp ridge structure in momentum space for linear and small elliptical polarization.
The final momenta of Coulomb-focused electrons that underwent slow recollisions, and hence experienced strong Coulomb focusing, are almost unaltered by the introduction of a small ellipticity. The ellipticity modifies the initial momenta of the electrons that end up in the ridge. We also learned that the central part of the tunnel ionized electron wave packet in momentum space provides electrons which are steered away by the elliptical polarization into elliptical side lobes of the final PMD.
Thus we were able to separate the unfocused from the Coulomb-focused electrons.      

Dispersion of the electron trajectories in final momentum was already reported in \cite{Shafir_2013}. However, there was no clear separation of the electrons undergoing Coulomb-focusing and the ones that interact only weakly with the Coulomb-field of the parent ion in momentum space.  
For small ellipticities we have shown that the electrons undergoing strong Coulomb-focusing accumulate in a sharp ridge in the 3D PMD that is largely separated from the unfocused electrons. This clear separation enables us to analyze non-dipole effects on Coulomb-focused electrons alone. 

We use the full 3D PMD to study the different non-dipole response for both electrons with and without strong Coulomb-focusing. 
The separation of the electron trajectories allows us to disentangle the response of the photoelectrons from non-dipole effects and to make the connection between prior results \cite{Ludwig_2014, Smeenk_2011}. In Ref. \cite{Smeenk_2011}, a shift of the PMD in beam propagation direction was reported and a simple radiation pressure picture was used as an explanation. In Ref. \cite{Ludwig_2014}, the peak of the projection of the PMD was shifted opposite to the beam propagation direction. Here, we were able to experimentally isolate the electrons that are responsible for this counterintuitive shift and to separately study their non-dipole response.      
We measured the ellipticity dependence of the non-dipole effects in PMDs in strong field ionization beyond the long-wavelength limit. We observed an offset of the maximum of the peak of the projection of the PMD on the beam propagation axis. By increasing the ellipticity from linear to circular, this offset shifts from negative to positive values of $p_z$. 
Thus, we conclude that the formation of Coulomb-focused structures in PMDs in combination with the magnetic field causes peaks in negative $p_z$-direction. The appearance of a peak on the negative or positive side of $p_z = 0$ can be considered as a competition between two types of photoelectrons which we refer to as type A and type B electrons (Fig. \ref{fig:Traj}). 
Type B photoelectrons form the sharp thin line-shaped ridge structure in the 3D PMD and type A photoelectrons are in the clearly separated lobes at higher final momentum. When we analyse the shift of the PMD peak along the laser beam propagation direction we could confirm that the type A electrons result in a positive shift and the type B electrons in a negative shift.
In case that the peak is formed mainly by type A-electrons, the projection of the PMD peaks at positive values of $p_z$ whereas in the case that the type B electrons dominate the peak of the projected PMD, it is shifted towards negative  values of $p_z$. 
Thus, Coulomb focusing has an essential influence on non-dipole effects and creates an ellipticity-dependent PMD shift along the laser propagation direction on the ellipticity of the laser field.

Our results open up new possibilities to separately study the photoelectrons experiencing strong rescattering from the direct electrons and the electrons that experience only weak Coulomb interaction. 
Furthermore, as the occurrence of the Coulomb focusing effects also depend on the width of the returning electron wave packet, our studies open up the possibilities to gain insight into the precise structure of the returning electron wave packet.

\appendix
\section{Derivation of the recollision conditions for elliptical polarization}
\label{sec:app}
We derive the condition under which the recollision parameters for elliptical polarization are similar to the ones for linear polarization. 

The laser field given by Eqs. (\ref{eq:laserField1}) and (\ref{eq:laserField2}), the electron relativistic equations of motion in the polarization plane of the laser field,  read: 
\begin{eqnarray}
\Lambda\omega\frac{dx}{d\eta} &=& -\frac{E_0}{\omega}\left(\sin \eta-\sin\eta_i\right)-\delta p_{xi}^C\label{dxdeta}\\
\Lambda\omega\frac{dy}{d\eta} &=&  \epsilon\frac{E_0}{\omega}\left(\cos \eta-\cos\eta_i\right)+p_{yi}-\delta p_{yi}^C,
\label{dydeta}
\end{eqnarray}
which is derived from Eqs. (\ref{px})-(\ref{elliptt-mom-y}), using the relation $\gamma d\eta/dt=\omega\gamma(1-\beta_z)\equiv \omega\Lambda$, where $\eta=\omega(t-z/c)$, $\Lambda$ is the integral of motion in a plane laser field \cite{RMP_2012}, $\beta_z$ and $\gamma$ are the electron beta- (along the laser propagation direction) and gamma-factors, respectively.
As the ionized electron appears at the tunnel exit with velocity much smaller than the speed of light, one has $\Lambda\approx 1$. The solution of Eqs. (\ref{dxdeta})-(\ref{dydeta}) is
\begin{eqnarray}
 x &=&  \frac{E_0}{\omega^2}\left(\cos \eta-\cos\eta_i\right)+\left[\frac{E_0}{\omega }\sin\eta_i-\delta p_{xi}^C\right]\frac{(\eta-\eta_i)}{\omega}+x_i,\label{x-ellipt}\\
 y &=&  \epsilon\frac{E_0}{\omega^2}\left(\sin \eta-\sin\eta_i\right)+\left[p_{yi}-\delta p_{yi}^C-\epsilon\frac{E_0}{\omega }\cos\eta_i \right]\frac{(\eta-\eta_i)}{\omega} +y_i,\nonumber \\\label{y-ellipt}
\end{eqnarray}
where the initial coordinates at the ionization phase $\eta_i$ correspond to the tunnel exit:
\begin{eqnarray}
x_i&=& x(\eta_i)=-\frac{I_pE_x(\eta_i)}{E^2(\eta_i)}=-\frac{I_p}{E_0}\frac{\cos \eta_i}{\cos^2 \eta_i+\epsilon \sin^2\eta_i},\\
y_i&=&y(\eta_i)=-\frac{I_pE_y(\eta_i)}{E^2(\eta_i)}= -\frac{I_p}{E_0}\frac{\epsilon\sin \eta_i}{\cos^2 \eta_i+\epsilon \sin^2\eta_i}.
\end{eqnarray}

Considering the electrons contributing to the ridge in the case of linear polarization, the recollision $y_r$-coordinate of the electron, ionized with the momentum $p_{yi}^{(0)}$, is 
\begin{eqnarray}
y_{r}^{(0)}=[p_{yi}^{(0)}-\delta p_{yi}^{C(0)}](t_r-t_i).
\label{yr0}
\end{eqnarray}
The momentum transfer upon recollision due to the Coulomb field is $\delta p_{yr}^{C(0)}$, and the final momentum of the electron is vanishing in the  case of linear polarization 
\begin{eqnarray}
p_{yf}^{(0)}=p_{yi}^{(0)}-\delta p_{yi}^{C(0)}-\delta p_{yr}^{C(0)}=0
\end{eqnarray}
Consequently, $p_{yi}^{(0)}=\delta p_{yi}^{C(0)}+\delta p_{yr}^{C(0)}$, and 
\begin{eqnarray}
y_{r}^{(0)}= \delta p_{yr}^{C(0)}(t_r-t_i).
\end{eqnarray}
According to Eq.~(\ref{y-ellipt}), the electron recollision coordinate in the case of elliptical polarization is 
\begin{eqnarray}
  y^{(\epsilon)}_r &=& \epsilon\frac{E_0}{\omega^2}\left(\sin \eta_r-\sin\eta_i\right)\label{yr} \\
  &+&\left[p_{yi}^{(\epsilon)}-\delta p_{yi}^{C(\epsilon)}-\epsilon\frac{E_0}{\omega }\cos\eta_i \right]\frac{(\eta_r-\eta_i)}{\omega} ,\nonumber 
 \end{eqnarray}
where $\delta p_{yi}^{C(\epsilon)}$ is the initial Coulomb momentum transfer in the case of elliptical polarization and where we set $y_i = 0$. The momentum transfer during recollision in the  case of elliptical polarization will be the same as in the case of linear polarization, $\delta p_{yr}^{C(\epsilon)}=\delta p_{yr}^{C(0)}\equiv \delta p_{yr}^{C}$,  if the impact parameter is the same, i.e., if $y_{r}^{(\epsilon)}=y_{r}^{(0)}$. The latter reads using Eqs. (\ref{yr}) and (\ref{yr0})
\begin{eqnarray}
\epsilon\frac{E_0}{\omega^2}\left(\sin \eta_r-\sin\eta_i\right)&+&\left[p_{yi}^{(\epsilon)}-\delta p_{yi}^{C(\epsilon)}-\epsilon\frac{E_0}{\omega }\cos\eta_i \right]\frac{(\eta_r-\eta_i)}{\omega} \nonumber \\
&\approx &\left[p_{yi}^{(0)}-\delta p_{yi}^{C(0)}\right]\frac{(\eta_r-\eta_i)}{\omega},
\end{eqnarray}
where we have assumed that $(z_r-z_i)\ll c(t_r-t_i)$.

Taking into account that for the slow recollision (points B2 and B4 in Fig. \ref{fig:01-comparison_of_PESs}) $p_{xr}=0$, which according Eq.~(\ref{px}) reads
\begin{eqnarray}
\frac{E_0}{\omega}\left(\sin \eta_r-\sin\eta_i\right)=-\delta p_{xi}^{C(\epsilon)},
\end{eqnarray}
we find that Coulomb focusing at the slow recollision in an elliptically polarized laser field is similar to the case of linear polarization if  
\begin{eqnarray}
 p_{yi}^{(\epsilon)}  =p_{yi}^{(0)}+\epsilon\frac{E_0}{\omega }\cos\eta_i +\delta p_{yi}^{C(\epsilon)}-\delta p_{yi}^{C(0)}+\frac{\epsilon\delta p_{xi}^{C(\epsilon)}}{\eta_r-\eta_i} .
\end{eqnarray}

\section{Derivation of the diameter and position of the initial momenta for the ridge}
\label{sec:app2}

We derive the equations for the center and the diameter of the ring of initial momenta contributing to a point on the ridge with $p_{yf} = 0$. Let $p_{yf}^{\pm}$ be the final momentum of an electron that started with $p_{zi} = 0$  from the left $(-)$ and the right $(+)$ part of the ring in initial momentum space, $p_{yi}^{+}>p_{yi}^{-} $, respectively. We begin with the condition:
\begin{eqnarray}
p_{yf}^-=p_{yf}^+,\label{==}
\end{eqnarray}
with
\begin{eqnarray}
p_{yf}^\pm=-\epsilon\frac{E_0}{\omega}\cos\eta_i+p_{yi}^\pm-\delta p_{yi}^{C\pm}\mp\delta p_{yr}^{C\pm}.
\end{eqnarray}
Using the expression for initial Coulomb momentum transfer: $\delta p_{yi}^{C }\approx2p_{yi} E(\eta_i)/(2I_p)^2$ \cite{Liu_2010}, $p_{yf}^\pm=0$, and the approximation $(2I_p)^2 \gg 2E(\eta_i)$ the initial momenta are given by
\begin{eqnarray}
p_{yi}^\pm=\left(\epsilon\frac{E_0}{\omega}\cos\eta_i\pm\delta p_{yr}^{C\pm}\right) \left(1+\frac{2E(\eta_i)}{(2I_p)^2} \right). 
\end{eqnarray} 
From Eq.~(\ref{==}) follows that the diameter $D$ of the ring in the initial momentum distribution depends linearly on Coulomb momentum
transfer during recollision:
\begin{eqnarray}
D\equiv p_{yi}^+-p_{yi}^-=\left(\delta p_{yr}^{C+} + \delta p_{yr}^{C-}\right)\left(1+\frac{2E(\eta_i)}{(2I_p)^2} \right).\label{DDD}
\end{eqnarray}
The center of the ring is dependent on the ellipticity $\epsilon$,
\begin{eqnarray}
\frac{p_{yi}^++p_{yi}^- }{2} = \left(\epsilon\frac{E_0}{\omega}\cos\eta_i+\frac{\delta p_{yr}^{C+}}{2}-\frac{\delta p_{yr}^{C-}}{2} \right) \left(1+\frac{2E(\eta_i)}{(2I_p)^2} \right),
\end{eqnarray}
and shifts to the right for increasing $\epsilon$.

\section{Derivation of the conditions for initial phase}
\label{sec:app3}
We can estimate the threshold of the ionization phase to have similarity of the recollision coordinate for linear and elliptical polarization, from the following condition: 
\begin{eqnarray}
|y_r^{(\epsilon)}-y_r^{(0)}|\ll y_r^{(0)},
\end{eqnarray}
with $y_r^{(\epsilon)}$ being the recollision $y-$coordinate at ellipticity $\epsilon$.
Using Eq.~(\ref{y-ellipt}) and using the approximations $\eta_r^{(\epsilon)} = \eta_r^{(0)}$ and $\eta_i^{(\epsilon)} = \eta_i^{(0)}$,
this condition reads
\begin{eqnarray}
 \left| \epsilon\frac{E_0}{\omega^2}\left(\sin \eta_r-\sin\eta_i\right)\right|\ll \left( p_{yi}^{(0)}-\delta p_{yi}^{C(0)} \right) \frac{(\eta_r-\eta_i)}{\omega},
\end{eqnarray}
which can be expressed via the Coulomb momentum transfer as:
\begin{eqnarray}
 \left|  \sin \eta_r-\sin\eta_i \right|\ll \frac{\delta p_{yr}^{C}}{(\epsilon E_0/\omega)} (\eta_r-\eta_i).
 \label{conditionetai}
\end{eqnarray}
The recollision phase is found from the condition $x (\eta_r)=0$. Taking into account that for the first recollision $\eta_r\approx 2\pi-\eta'_r$, with $\eta'_r\ll 1$, we find 
\begin{eqnarray}
\eta'_r\approx \sqrt{4\pi\eta_i},
\end{eqnarray}
and the condition of Eq.~({\ref{conditionetai}}), when the oscillating part of the $y$-coordinate does not perturb the recollision coordinates significantly, is
\begin{eqnarray}
\eta_i\ll \pi \left(\frac{\delta p_{y}^{C}}{(\epsilon E_0/\omega)} \right)^2.
\end{eqnarray}
For $\delta p_{y}^{C} = 0.1$, we have $\eta_i \ll  0.68$.

\begin{acknowledgments}
This research was supported by the NCCR MUST, funded by the Swiss National Science Foundation and by the ERC advanced grant ERC-2012-ADG 20120216 within the seventh framework programme of the European Union. Benjamin Willenberg was supported by an ETH Research Grant ETH-11\~15-1.
\end{acknowledgments}
 
\bibliography{LineResSelected}

\begin{thebibliography}{70}
\bibitem [{Tun()}]{TunnelIonization}%
  \BibitemOpen
  \href@noop {} {}\bibinfo {howpublished} {Here, we refer to the theory of
  tunnel ionization in the quasistatic limit. Commonly, this case is reffered
  to as the Ammosov-Delone-Kraĭnov (ADK). The results of this theory are
  already contained in the Perelomov-Popov-Terent'ev (PPT)-theory as the limit
  for the a zero-frequency field.}\BibitemShut {Stop}%
\end{thebibliography}

\end{document}